\documentclass[runningheads]{llncs}
\usepackage{hyperref}  
\usepackage{graphicx}
\begin{document}
\title{Adapting LLMs for Efficient, Personalized Information Retrieval: Methods and Implications}
%
%
\author{Samira Ghodratnama \and
Mehrdad Zakershahrak}
\authorrunning{S. Ghodratnama et al.}
%
\institute{Macquaire University, Sydney, Australia \\
\email{\{samira.ghodratnama,mehrdad.zakershahrak\}@mq.edu.au}}
\maketitle              
\begin{abstract}
The advent of Large Language Models (LLMs) heralds a pivotal shift in online user interactions with information.
Traditional Information Retrieval (IR) systems primarily relied on query-document matching, whereas LLMs excel in comprehending and generating human-like text, thereby enriching the IR experience significantly.
While LLMs are often associated with chatbot functionalities, this paper extends the discussion to their explicit application in information retrieval.
We explore methodologies to optimize the retrieval process, select optimal models, and effectively scale and orchestrate LLMs, aiming for cost-efficiency and enhanced result accuracy.
A notable challenge, model hallucination—where the model yields inaccurate or misinterpreted data—is addressed alongside other model-specific hurdles.
Our discourse extends to crucial considerations including user privacy, data optimization, and the necessity for system clarity and interpretability.
Through a comprehensive examination, we unveil not only innovative strategies for integrating Language Models (LLMs) with Information Retrieval (IR) systems, but also the consequential considerations that underline the need for a balanced approach aligned with user-centric principles.

\end{abstract}
\section{Introduction}
In the current digital landscape, there's a massive increase in data.
This has intensified the need for systems that can effectively retrieve the right information.
Information retrieval plays a crucial role in helping us navigate the vast amount of digital data, requiring methods that are both accurate and sensitive to context.
The systems should not only be precise but also capable of understanding the context to provide users with the most relevant results.
Additionally, they should be adept at summarization to condense information, thereby enabling users to quickly grasp the essence of the content and make well-informed decisions~\cite{ghodratnama2020extractive,ghodratnama2020rare,ghodratnama2020adaptive}.

While traditional information retrieval systems have been helpful, they sometimes struggle to fully grasp the context of queries~\cite{beheshti2021query}.
This can lead to results that don't quite match what the user was looking for~\cite{ghodratnama2021summary2vec}.
Recent advances have brought forward neural network models, which have seen many improvements over the years~\cite{khanna2022transformer}.
Among these, Large Language Models (LLMs) stand out.
They have exhibited notable capabilities in generating human-like language across various applications, from executing language tasks to creating content.
They can understand and create text that's very similar to how humans write, due to the extensive data they've been trained on. 

Despite their ability to produce seemingly fluent and authoritative responses, LLMs also present risks, making them principally advantageous for prototyping while often being limited in broader production contexts.
One significant risk is the production of inaccurate or "hallucinated" information, in addition to confronting challenges related to bias, consent, and security.
Secondly, their knowledge is confined to the information available up to the point of their last training, restricting their ability to utilize new data.
Thirdly, their effectiveness in specialized tasks can be limited due to their generalized training background. Consequently, in a business context, granting excessive control to an LLM may yield unwelcome results, prompting it to generate irrelevant, harmful, or even dangerous content, all while exuding a misleadingly high level of confidence. 
Furthermore, the frequent refinement of these models can incur substantial costs.
Given these risks, a pivotal question arises: How can we wisely employ the strengths of LLMs in our product development while minimizing potential drawbacks? 
It is imperative to acknowledge and navigate their inherent limitations, applying meticulous evaluation and probing techniques for specific applications, rather than solely relying on ideal-case interactions.

In this paper, we navigate through the extant challenges linked with Large Language Models (LLMs), addressing the pivotal question: `Can integrating LLMs with current retrieval technologies engender a novel, enhanced strategy for information retrieval?'
We explore potential solutions and scrutinize a variety of extant approaches. Our discourse is anchored steadfastly in the pragmatic aspects of the challenge, aiming to ensure that our findings and recommendations reverberate meaningfully across both academic and industrial domains.
Our objective is to enrich the discourse in the field of information retrieval  and delineate a trajectory for impending advancements in the field.

\section{Related work}
The quest for better ways to extract information from large data repositories has been a constant theme in computational linguistics and computer science, especially with the complexity of the Internet's hyperlink networks~\cite{beheshti2022social}.
Algorithms like PageRank~\cite{pagerank} have tackled this challenge, navigating this network to evaluate the importance of web pages and bring some order to the vastness of the Internet.
Early on, notable work by researchers like Salton and McGill introduced the Vector Space Model (VSM)~\cite{salton1975vector} to the global academic and scientific community, setting the stage for many advancements to come.
This model conceptualized text documents as vectors within a multi-dimensional space, igniting curiosity that subsequently catalyzed the advent of neural network-based language models~\cite{ghodratnama2021towards,ghodratnama2023personalized,ghodratnama2021intelligent}.
The journey from early sequence modeling, represented by RNNs and LSTMs~\cite{hochreiter1997long}, to the transformative era brought about by transformer models~\cite{vaswani2017attention}, has been swift. 
Large Language Models (LLMs) like OpenAI's GPT series and Google's BERT have gone beyond just text retrieval or recognition, evolving to understand and generate text in a way that is much like human communication~\cite{radford2019language,devlin2018bert,zakershahrak2020we}.

In the sections that follow, we delve into the crucial role of LLMs in enhancing information retrieval, with a special focus on employing Retrieval-Augmented Generation (RAG)~\cite{lewis2020retrieval} to address the prevalent challenges associated with language models.
We outline the architecture of RAG models and discuss their applicability and efficacy in various settings. Furthermore, we explore the integration of knowledge bases with information retrieval systems to augment the richness of contextual understanding and provide a more comprehensive and accurate retrieval of information. 
This integration is pivotal in bridging the gap between structured and unstructured data, thereby facilitating a more informed and insightful interaction for users~\cite{processGPT,AminICWS}.

\section{LLM as IR Enhancer: Methods and Implications}
Despite the remarkable capabilities of foundation models, their utilization as direct information sources presents challenges.
In this section, we expound upon key discussions related to employing language models within the context of information retrieval.
We navigate through prevailing challenges, explore potential solutions applicable in varied situations.
We particularly discuss the following considerations in context of information retrieval:
\begin{itemize}
    \item Is fine-tuning a model advisable in an enterprise setting?
    \item What factors should be scrutinized before deploying large language models?
    \item How can different components be effectively integrated?
    \item What metrics should be in place for continuous performance evaluation?
    \item What security measures and privacy compliance standards need to be adhered to?
\end{itemize}

\begin{figure*}[t]
    \centerline{\includegraphics[width=\textwidth]{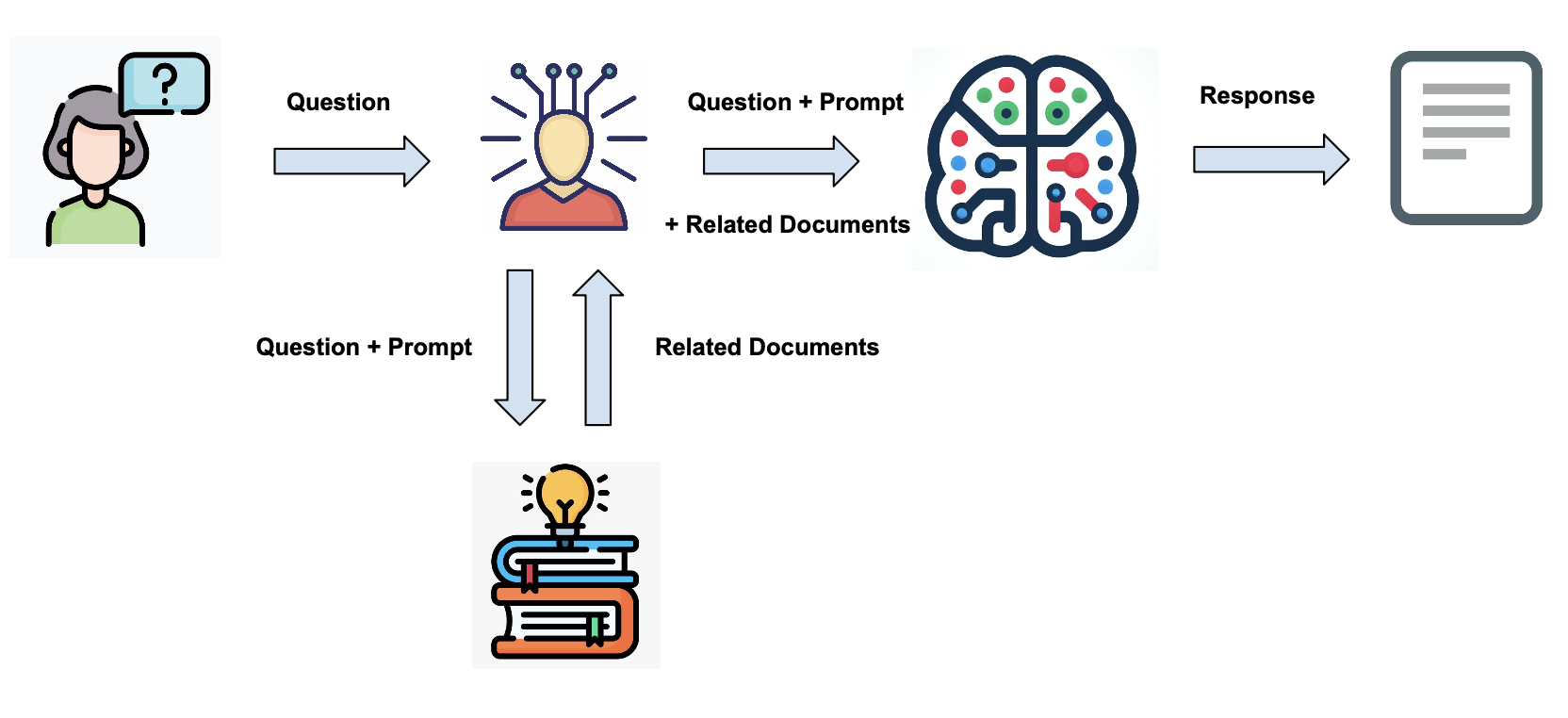}}
    \caption{An overview of the RAG architecture.}
    \label{rag}
\end{figure*}
\subsection{Enhancing the Information Retrieval Process}
Foundation models, although powerful, encounter several limitations due to their intrinsic reliance on static, pre-trained knowledge, hampering their capability to stay abreast of contemporary developments and demonstrating difficulties in executing specialized, domain-specific tasks.
Moreover, these models, while being vast reservoirs of general knowledge, sometimes fail to navigate the intricacies and depth required in specialized domains.
This issue, paired with the risk of ``hallucination" – generating plausible but incorrect or nonsensical information – underscores the importance of seeking enhancements to leverage these models effectively and safely.

Retrieval-Augmented Generation (RAG) emerges as a pivotal technology to address some of these limitations, forging a connection between the foundational knowledge of large language models and the dynamic, up-to-date information contained in external repositories.
Imagine a system that goes beyond just understanding a query, having the intelligence to know where to look for answers and the ability to form responses coherently and contextually.
This vision, named Retrieval-Augmented Generation (RAG), was detailed by Lewis et al.~\cite{lewis2020retrieval}, showing a close connection between retrieval systems and the generative capabilities of language models.
An overview of teh RAG achitecture id depicted in Figure~\ref{rag}.
This approach strengthens the model's ability to find relevant information and present it in a useful and contextually relevant way, enhancing the usefulness of machine learning models in real-world information query scenarios.
By facilitating real-time retrieval of relevant information during the inference process, RAG circumvents the static knowledge limitation, providing outputs that are not only rich and contextually relevant but also verifiable against a known dataset.

Navigating the practical implementation of RAG brings several key advantages to the forefront.
The potential for cost and computational savings is evident, as the need for frequent, exhaustive retraining of the model is mitigated, supplanted by updates to the retrieval database.
Interestingly, the approach of RAG also lends a hand in addressing privacy concerns.
Since RAG pulls from an external retrieval database, instead of integrating data into the model, it aids in keeping sensitive information more secure. 
This characteristic ensures that the responses generated are current and relevant, while also providing a structured way to handle sensitive or private information within the retrieval data.
This setup aims at striking a careful balance between dynamism and privacy.

A question that often arises is whether RAG can be replaced by fine-tuning. 
The distinction between RAG and fine-tuning becomes apparent when considering the dynamic nature of the information involved, the computational resources available for retraining, and the specific requirements of the task at hand.
While both methodologies aim at enhancing the performance and applicability of Large Language Models (LLMs), they cater to different aspects and scenarios.
Fine-tuning is a well-suited approach for tailoring a model to a specific domain or set of long-term tasks, especially when the underlying challenges are relatively static.
It involves adjusting the model parameters on a new dataset to make the model's behavior more aligned with the desired task or domain.
On the other hand, RAG is designed to tackle scenarios where the model needs to stay updated with rapidly evolving or expansive information without requiring continuous retraining.
By leveraging an external retrieval database, RAG enables the model to interact dynamically with the latest data, ensuring its responses are current and relevant.
While fine-tuning demands substantial computational resources for retraining with new data, RAG offers a cost-effective alternative by minimizing the need for exhaustive retraining.
In summary, the choice between RAG and fine-tuning is not a matter of simple replacement, but rather a strategic decision based on the particular demands of the task, the nature of the data, and the available resources. 
Each approach has its own set of advantages and is suited to different use cases, necessitating a careful consideration of the project's requirements to determine the most appropriate strategy.

\subsection{Determining the Optimal Model: Criteria and Considerations}
Selecting a suitable Large Language Model (LLM) requires a careful look at different factors.
Domain specificity is crucial.
Specialized models often perform better than general-purpose ones in specific areas.
Given the variety in their training datasets, different models may be more suitable, each aligning well with different scenarios in their own way.

Scalability is a key feature when selecting an LLM.
This refers to the model's ability to effectively manage increased demands, handle a growing number of requests without a significant drop in performance.
Computational costs, infrastructure needs, the possible need for distributed processing, and a smooth transition to updated models or versions are factors that highlight an LLM's scalable nature.
A truly scalable LLM should not only meet immediate needs but also anticipate and accommodate future growth, ensuring continued relevance and performance in a rapidly changing environment.

Addressing bias and fairness is critical when deploying LLMs.
LLMs, trained on vast web-based data, may unintentionally reflect and spread societal biases present in their training data, potentially producing outputs that reinforce stereotypes or marginalize groups.
Moreover, how a model deals with biases and possibly produces inappropriate content, as well as its ease of integration and availability of support, tools, and community resources, all play a role in its usefulness and upkeep in practical situations.
Ensuring fairness requires that outputs are unbiased and don't unfairly favor or disadvantage any group, needing ongoing, use-case-specific testing, tuning, and validation.

Data privacy is also a major concern, requiring measures to ensure that LLMs neither retain nor disclose sensitive data and comply with global data protection standards.
Moreover, transparency and interpretability, although challenging due to the 'black-box' nature of LLMs, are essential for trust and understanding the reasoning behind the model’s outputs, requiring methods that reveal its workings to users and stakeholders.



\subsection{Orchestrating and Deploying LLM-based Applications}
Solving advanced tasks with language models and retrieval models  require frameworks to unify techniques for prompting and fine-tuning LMs — and approaches for reasoning, self-improvement, and augmentation with retrieval and tools. 
Langchain\footnote{https://www.langchain.com} serves as a pioneering framework tailored for crafting applications that harness the power of language models. 
At its heart, it operates on the principle that multiple components can be seamlessly chained together, enabling the design of sophisticated use cases centered around LLMs.
While Langchain offers a robust solution, the ecosystem is enriched by the presence of other notable frameworks.
LlamaIndex\footnote{https://www.llamaindex.ai} and DSPy\footnote{https://github.com/stanfordnlp/dspy}, for instance, also bring unique capabilities to the table, each contributing to the expansive landscape of language model application development.
Navigating through the domain of sophisticated language model application frameworks, it is pivotal to delve into the strengths and capabilities of the emerging libraries like Langchain, LlamaIndex, and DSPy.
These libraries proffer unique functionalities and methodologies, each converging on the development and deployment of Large Language Models (LLMs) but diverging in the nuances of their application and usability.

Langchain emerges as a front-runner, providing a pioneering framework that meticulously blends various components and tools to harness the might of LLMs, thereby establishing itself as a versatile framework, especially for developers seeking a flexible and extensible interface for a general-purpose application. 
Its principle revolves around the seamless chaining of multiple components, creating a vibrant ecosystem that is not only adaptable but can synergistically integrate with numerous external models, which in turn, pave the path for the deployment of innovative and versatile solutions.
The inherent flexibility of Langchain enables the realization of varied use-cases, thus broadening the horizons for developers engrossed in crafting applications anchored on LLMs.

On the flip side, LlamaIndex, while maintaining some functional overlap with Langchain, prioritizes efficiency and simplicity in search and retrieval applications, particularly via its conversational interface.
It's not merely a tool but an intuitive facilitator that empowers developers to efficiently manage, search, and summarize documents by utilizing LLMs and inventive indexing techniques, with graph indexes being a cornerstone feature.
It is interesting to observe how LlamaIndex and Langchain, while overlapping in functionalities such as data-augmented summarization and question answering, distinguish themselves in their approach and utility.
LlamaIndex’s intensive utilization of prompting and its prowess in creating hierarchical indexes for efficient data organization carve its unique niche in the landscape of language model application development. 
However, Langchain offers a more granular control and caters to an expansive array of use-cases.

On a slightly divergent trajectory, DSPy maneuvers the pathway by emphasizing programming, complemented by an encompassing approach that unifies techniques for not only prompting and fine-tuning LMs but also enhancing them through reasoning and tool/retrieval augmentation.
Its distinctive facet lies in its ability to provide composable and declarative modules for instructing LMs in a syntax familiar to developers acquainted with Python. 
Moreover, DSPy's automatic compiler, which traces programs and crafts premium prompts or trains automatic finetunes, exemplifies a nuanced approach to instructing LMs in task steps, thus presenting an alternative methodology compared to Langchain and LlamaIndex.

Although LlamaIndex and Langchain boast frequent updates and a steadfast evolution, posing a potential for amalgamation in the future, and DSPy treads its own unique path with a definitive emphasis on programming and a Pythonic approach, all three libraries together enrich the developer's toolkit, each offering varied approaches and functionalities catering to diverse application needs in the realm of language model application development and the choice between them would pivot on the specific requisites of the project.

\subsection{Examining Language Learning Model-Integrated Applications}
In traditional machine learning problems, 'Ground Truths' are pivotal for model building as they measure the quality of the model's predictions.
They are crucial for determining which model experiment should be deployed in production and help teams sample and annotate production data to identify and improve cohorts of low model performance.
Before the advent of Large Language Models, which excel in custom use-cases right out of the box, model evaluation followed a fairly straightforward protocol: data was split into training, test, and dev sets, with the model being trained on the training set and evaluated on the test/dev set.
This category also includes transfer learning and fine-tuning of models.
LLMs, however, navigate scenarios where establishing a clear Ground Truth is arduous.

Given the multi-dimensional nature of the problem, a comprehensive performance evaluation framework is necessitated.
Proper evaluation metrics should encompass all aspects of an application.
Key engineering aspects required for model performance evaluation during development include:
\begin{itemize}
    \item Prompt Tuning: Framing the prompt differently can yield considerably varied results.
    \item Embeddings Model: Connecting custom data with LLMs provides the necessary context. The choice of embeddings and similarity metrics is crucial.
    \item Model Parameters: Parameters like temperature, top-k, and repeat penalty in LLMs are noteworthy.
    \item Data Storage: For advanced retrieval augmented generation (RAG) solutions, considerations on how to store and retrieve data, such as storing document keywords alongside embeddings, are important.
\end{itemize}

There are some other aspects that need to be evaluated post-development, which include:
\begin{itemize}
    \item Accuracy: Are the answers relevant to the query?
    \item Speed/Response time: Is the answer generated within a reasonable timeframe?
    \item Completeness: Are all relevant items retrieved?
    \item Error rate: Frequency of errors or incorrect information.
    \item Prompt quality: The number of interactions needed to attain the desired result.
    \item Output structure: The quality of the presented answer.
\end{itemize}

With evaluation parameters in place, deciding on the evaluation metrics and model performance evaluation becomes feasible.
Generating evaluation datasets from representative user inputs might pose a challenge; however, benchmark datasets like MTBench, MMLU, and TruthfulQA can serve as guidelines.
Recent developments like RAGAS~\footnote{https://github.com/explodinggradients/ragas} and LangSmith~\footnote{https://www.langchain.com/langsmith} have attempted to streamline LLM evaluation based on various metrics such as faithfulness, relevancy, and harmfulness which provide a framework to debug, test, evaluate, and monitor LLM applications respectively.
The shift from employing Generative AI as personal search engines to integrating them into production underscores the exciting prospects in deploying, monitoring, and evaluating LLMs.

There is also a new line of research on how to evaluate LLM-based applications.
A recent paper by Zheng et al.~\cite{zheng2023judging} proposed employing robust LLMs as judges to evaluate applications on nuanced, open-ended questions. 
By introducing two benchmarks, MT-bench and Chatbot Arena, it aims to bridge the conventional evaluation metrics and human preferences gap. 
The paper explores the potential of using state-of-the-art LLMs like GPT-4 as a surrogate for human judges, termedc``LLM-as-a-judge," to automate the evaluation process.
A systematic study revealed that the LLM-as-a-judge approach could achieve over 80\% agreement with human evaluations, matching the level of human-human agreement.
This suggests a scalable and swift method to evaluate human preference in LLM-based applications, presenting a promising alternative to traditionally slow and costly human evaluations.
The study also emphasizes the importance of a hybrid evaluation framework, amalgamating existing capability-based benchmarks and new preference-based benchmarks with the LLM-as-a-judge approach, for a more comprehensive evaluation of both core capabilities and human alignment of models.

\section{Conclusion}
In conclusion, this paper sheds light on the nuanced advantages and challenges tied to employing Large Language Models (LLMs) in refining Information Retrieval (IR) systems.
Emphasizing personalized and efficient information retrieval, we investigated various strategies to bolster the retrieval process, choose suitable models, scale and manage LLMs effectively, and thoroughly assess their performance and impacts. 
LLMs, with their capacity for human-like text comprehension and generation, extend beyond traditional query-document matching, paving the way for a richer, personalized IR experience.

Nevertheless, the path to fully unlocking LLMs' potential in IR systems is laden with obstacles. 
Notable challenges include model hallucination, ensuring user privacy, optimizing data usage, and maintaining system clarity and interpretability.
These issues underscore the necessity for a balanced approach encompassing technical innovations, ethical considerations, and user-centric design.

The dialogue advanced in this paper advocates for a harmonious fusion of LLMs and IR systems, inviting a collective effort among researchers, practitioners, and policymakers to tackle model-centric challenges and broader ramifications.
As the lines between human and machine interaction blur further, the collaborative intelligence of this human-machine amalgam is instrumental in nurturing an IR ecosystem that is efficient, personalized, and accountable.
Through relentless exploration and iterative improvements, the potential of LLMs in substantially elevating the IR domain is not only conceivable but also promising.


\bibliographystyle{elsarticle-num}
\bibliography{main.bib}
\end{document}